\documentclass{iopart}
\usepackage{graphicx}
\usepackage{cite}
\begin{document}

\paper{Dynamics at a smeared phase transition}

\author{Bernard Fendler$^{1,2}$, Rastko Sknepnek$^{1,3}$ and Thomas Vojta$^1$}

\address{$^1$Department of Physics, University of Missouri - Rolla, Rolla, MO 65409, USA}
\address{$^2$Department of Physics, Florida State University, Tallahassee, FL 32306, USA}
\address{$^3$Department of Physics and Astronomy, McMaster University, Hamilton ON L8S 4M1, Canada}

\begin{abstract}
We investigate the effects of rare regions on the dynamics of Ising magnets with planar
defects, i.e., disorder perfectly correlated in two dimensions. In these systems, the
magnetic phase transition is smeared because static long-range order can develop on
isolated rare regions. We first study an infinite-range model by numerically solving
local dynamic mean-field equations. Then we use extremal statistics and scaling arguments to
discuss the dynamics beyond mean-field theory. In the tail region of the smeared
transition the dynamics is even slower than in a conventional Griffiths phase: the spin
autocorrelation function decays like a stretched exponential at intermediate times before
approaching the exponentially small equilibrium value following a power law at late times.

\end{abstract}


\submitto{\JPA}

\section{Introduction}
\label{sec:intro}

In recent years, the influence of quenched disorder on phase transitions and critical
points has reattracted considerable attention. A particularly interesting source of new
phenomena are the effects of rare spatial regions which can be locally in the wrong
phase. These rare regions or "Griffiths islands" fluctuate very slowly because flipping
them requires changing the order parameter in a large volume. Griffiths \cite{Griffiths}
showed that this leads to a non-analytic free energy everywhere in an entire temperature
region close to the transition, which is now known as the Griffiths phase \cite{Randeria}
or the Griffiths region. In generic classical systems with uncorrelated disorder, the
contribution of the Griffiths singularities to thermodynamic (equilibrium) observables is
very weak since the singularity in the free energy is only an essential one
\cite{Griffiths,Bray89}. In contrast, the consequences for the dynamics are much more
severe with the rare regions dominating the behavior for long times. In the Griffiths
region, the spin autocorrelation function $C(t)$ decays very slowly with time $t$, like
$\ln C(t) \sim -(\ln t)^{d/(d-1)}$ for Ising systems
\cite{Randeria,Dhar,Dhar88,Bray88a,Bray88b}, and like $\ln C(t) \sim -t^{1/2}$ for
Heisenberg systems \cite{Bray88a,Bray87}. Many of these results have been confirmed by
rigorous calculations of disordered Ising models \cite{Dreyfus,Gielis95,Cesi97}.

The effects of impurities are greatly enhanced by long-range spatial disorder
correlations. In an Ising model with linear defects, the thermodynamic Griffiths
singularities are given by power-laws with the average susceptibility actually diverging
in a finite temperature region. The critical point itself is of exotic
infinite-randomness type and displays activated rather than power-law scaling. This was
first found in the McCoy-Wu model \cite{McCoyWu,McCoy69}, a disordered 2D Ising model in
which the disorder is perfectly correlated in one dimension and uncorrelated in the
other. Later it was studied in great detail in the context of the quantum phase
transition of the random transverse-field Ising model where the imaginary time dimensions
plays the role of the "correlated" direction \cite{dsf9295}.

Recently, it has been shown that even stronger effects than these power-law (quantum)
Griffiths singularities can occur in Ising models with planar defects
\cite{jpa2003,prb2004}. Because the disorder is perfectly correlated in two directions,
the effective dimensionality of the rare regions is two. Therefore, an isolated rare
region can undergo the phase transition independently from the bulk system. This leads to
a destruction of the global sharp phase transition by smearing. Similar smearing effects
have also been found in itinerant quantum magnets \cite{rounding_prl} and at a
non-equilibrium transition in the presence of linear defects \cite{contact_pre}. In all
of these cases, the effective dimensionality of a rare region is above the lower critical
dimension of the problem which allows the rare region to order independently.

In this paper, we investigate the {\em dynamics} of an Ising model with planar defects in
the vicinity of this smeared phase transition. We show that the dynamics in
the "tail" of the smeared transition is even slower than in a conventional Griffiths phase.
The autocorrelation function decays like a
stretched exponential at intermediate times followed by power-law behavior at late times.
The paper is organized as follows. In section \ref{sec:td}, we introduce the model and
briefly summarize the results for the thermodynamics of the smeared
transition to the extent necessary for the discussion of the dynamics. In section
\ref{sec:numerics}, we numerically solve dynamical mean-field equations and compare the
results to predictions based on a variational approach. In section \ref{sec:dyn}, we use
general scaling arguments in connection with extremal statistics to go beyond mean-field
theory.  Conclusions are presented in section \ref{sec:conclusions}.

\section{The model and its thermodynamics} \label{sec:td}

For definiteness we consider a $d$-dimensional Ising model with bond disorder completely
correlated in the $d_C=2$ directions $x_1$ and $x_2$ but uncorrelated in the remaining
$d_\bot=d-d_C$ directions $x_3,\dots,x_d$. Classical Ising spins $S_{\bf x}=\pm 1$ reside
on the sites ${\bf x}$ of a hypercubic lattice. They interact via nearest-neighbor
interactions. In the clean system all interactions are independent of the lattice site;
their values are $J_0$ for bonds in the uncorrelated directions and $J_0/4$ for bonds in
the $x_1$ and $x_2$ directions. We have chosen this scaling of the interactions because
it leads to mean-field equations compatible with the corresponding thermodynamic
mean-field theory \cite{jpa2003}. The planar defects are modelled by weakening the bonds
in ($x_1,x_2$)-planes at random positions in the uncorrelated directions. The system
effectively consists of blocks separated by parallel planes of weak bonds. The
Hamiltonian of the system is given by:
\begin{eqnarray}
\label{eq:Hamiltonian} H=&&-\frac 1 4 \sum_{\bf x}  J({\bf x}_\bot) ~\left [ S_{{\bf x}}
S_{{\bf x}+{\bf e_1}} + S_{{\bf x}} S_{{\bf x}+{\bf e_2}} \right ] -\sum_{\bf x}
\sum_{j=3}^d J_0 ~S_{{\bf x}} S_{{\bf x}+{\bf e_j}},
\end{eqnarray}
where ${\bf e}_j$ is a unit vector in direction $x_j$, ${\bf x}_\bot$ is the projection
of ${\bf x}$ on the uncorrelated directions $x_3,\dots,x_d$, and $J({\bf x}_\bot)$ is the
random coupling constant in the ($x_1,x_2$)-planes. The values of $J({\bf x}_\bot)$ are
drawn from a binary distribution:
\begin{equation}
\label{eq:2} J({\bf x}_\bot)=\left\{
\begin{array}{cl}
 cJ_{0} & \textrm{with probability}~ p\\ J_{0}  & \textrm{with probability}~ (1-p)
\end{array}
\right .
\end{equation}
characterized by the concentration $p$ and the relative strength $c$ of the weak bonds
($0\le p\le 1, 0<c\le 1$).

We now briefly summarize the results on the thermodynamics of this system \cite{jpa2003,prb2004}
to the extent necessary for the discussion of the dynamics. The clean system ($p=0$) undergoes
a magnetic phase transition at some temperature $T_c^0$, with the order parameter being the total magnetization
\begin{equation}
\label{eq:3} m=\frac{1}{V} \sum_{{\bf x}} \langle S_{{\bf x}} \rangle,
\end{equation}
where $V$ is the system volume, and $\langle\cdots\rangle$ is the thermodynamic average.
In the disordered system ($p>0$), a crucial role is played by rare strong disorder
fluctuations: There is a small but non-zero probability for finding large spatial regions
in $\bf x_\bot$ direction devoid of any impurities. These regions can be locally in the
ordered phase even if the bulk system is in the disordered phase. Because of the disorder
correlations, these rare regions are infinite in the $x_1$ and $x_2$ directions but
finite in the remaining directions. Thus, each rare region is equivalent to a
two-dimensional Ising system and can undergo a real phase transition {\em independently}
of the rest of the system. The resulting effect is much stronger than the conventional
Griffiths effects \cite{Griffiths,Randeria}: The global phase transition is destroyed by
smearing \cite{jpa2003}; and  the order parameter develops very inhomogeneously in space
with different parts of the system (i.e., different ${\bf x}_\bot$ regions) ordering
independently at different temperatures. Correspondingly, the correlation length in ${\bf
x}_\bot$ direction remains finite for all temperatures.

Using extremal statistics, the leading thermodynamic behavior can be easily worked out in
the 'tail' of the smeared transition, i.e., in the parameter region where a few islands
have developed static order but their density is so small that they can be treated as
independent. The probability $w$ of finding a large region of linear size $L_R$ (in ${\bf
x}_\bot$-space) devoid of any impurities is, up to pre-exponential factors, given by
\begin{equation}
w\sim (1-p)^{L_R^{d_\bot}} =  \exp[  L_R^{d_\bot} \ln(1-p)]~. \label{eq:wLR}
\end{equation}
As discussed above, such a rare region develops static long-range (ferromagnetic) order at
some temperature $T_c(L_R)$ below the clean bulk critical temperature $T_c^0$. The value of
$T_c(L_R)$ varies with the size of the rare region; the largest islands will develop
long-rage order closest to the clean critical point. According to finite size scaling we
obtain
\begin{equation}
\label{eq:FSS} T_c(L)-T_c^0=r_c(L_R)=-AL_R^{-\phi},
\end{equation}
where $\phi$ is the finite-size scaling shift exponent of the clean system and $A$ is the
amplitude for the crossover from $d$ dimensions to a slab geometry infinite in two
dimensions but finite in $d_\bot=d-2$ dimensions. The reduced temperature $r = T -T_c^0$
measures the distance form the \emph{clean} critical point. If hyperscaling is valid, the
finite-size shift exponent fulfills $\phi=1/\nu$.
 Combining (\ref{eq:wLR}) and (\ref{eq:FSS}) we obtain the
probability for finding a rare region that becomes critical at $r_c$ as
\begin{equation}
w(r_c) \sim \exp  (-B ~|r_c|^{-d_\bot/\phi}) \qquad (\textrm{for } r_c\to 0-)
 \label{eq:w-dilute}
\end{equation}
where the constant $B$ is given by $B=-\log(1-p)\,A^{d_\bot/\phi}$.  The total (or
average) order parameter $m$ is obtained by integrating over all rare regions which are
ordered at $r$, i.e., all rare regions having $r_c>r$. Since the functional dependence on
$r$ of the order parameter on a given island is of power-law type it does not enter the
leading exponentials but only the pre-exponential factors. To exponential accuracy, we
therefore obtain
\begin{eqnarray}
m \sim \exp  (-B ~|r|^{-d_\bot/\phi})  \qquad~
(\textrm{for } r\to 0-)~. \label{eq:m-dilute}
\end{eqnarray}

 Thus, the total
magnetization develops an exponential tail towards the disordered phase which reaches all
the way to clean critical point. Analogous estimates show that the homogeneous magnetic
susceptibility does not diverge anywhere in the tail region of the smeared transition
because the exponentially decreasing island density overcomes the power-law divergence of
the susceptibility of an individual island. However, there is an essential singularity at
the clean critical point produced by the vanishing density of ordered islands.

\section{Local dynamic mean-field approach}
\label{sec:numerics}

We now turn to the main topic of this paper, \emph{viz.}, the dynamic behavior in the
vicinity of the smeared
phase transition. We consider a purely relaxational local dynamics without any conservation
laws, i.e., model A in the Hohenberg-Halperin classification \cite{hh1977}.
Microscopically, it can be realized, e.g., by the Glauber algorithm \cite{glauber}.

In this section, we study the dynamics of the Ising model (\ref{eq:Hamiltonian}) by
numerically solving local dynamic mean-field equations. For definiteness, we consider
$d=3$ spatial dimensions, i.e., $d_\bot=1$ and $d_C=2$. A local dynamic mean-field theory
of the Hamiltonian (\ref{eq:Hamiltonian}) can be derived, e.g., by following Ref.\
\cite{stanley}. The plane magnetizations
\begin{equation}
m_{x_3}= \frac 1 {L_C^2} \sum_{x_1,x_2} S_{\bf x}
\end{equation}
where $L_C$ is the system size in correlated direction, then fulfill the local mean-field
equations of motion
\begin{equation}
\frac 1 \alpha \frac d {dt} m_x = - m_x + \tanh \left[(J_0 m_{x-1} + J(x)m_x + J_0
m_{x+1})/T  \right]~. \label{eq:mf}
\end{equation}
Here, $\alpha$ is a microscopic relaxation rate which determines the overall time scale.
Alternatively, the same equations can be obtained without further approximation from the
infinite-range Hamiltonian used in the thermodynamic mean-field theory \cite{jpa2003}.

Note that the local mean-field equations contain the full spatial dependence of the
magnetization on the uncorrelated dimension. In contrast to a global mean-field theory
these local equations are therefore capable of describing rare region physics. The
stationary solution of (\ref{eq:mf}) corresponds to the thermodynamic local mean-field
theory studied in Ref.\ \cite{jpa2003}. The clean model ($p=0$) undergoes a magnetic
phase transition at $T_c^0=3 J_0$, displaying mean-field critical behavior. Below
$T_c^0$, the diluted system ($p>0$) develops an exponential magnetization tail of the
form $m \sim \exp(-B|T_c^0-T|^{-1/2})$ following the prediction (\ref{eq:m-dilute}) with
$d_\bot=1$ and $\phi=2$.

\begin{figure}
\includegraphics[width=0.5\textwidth]{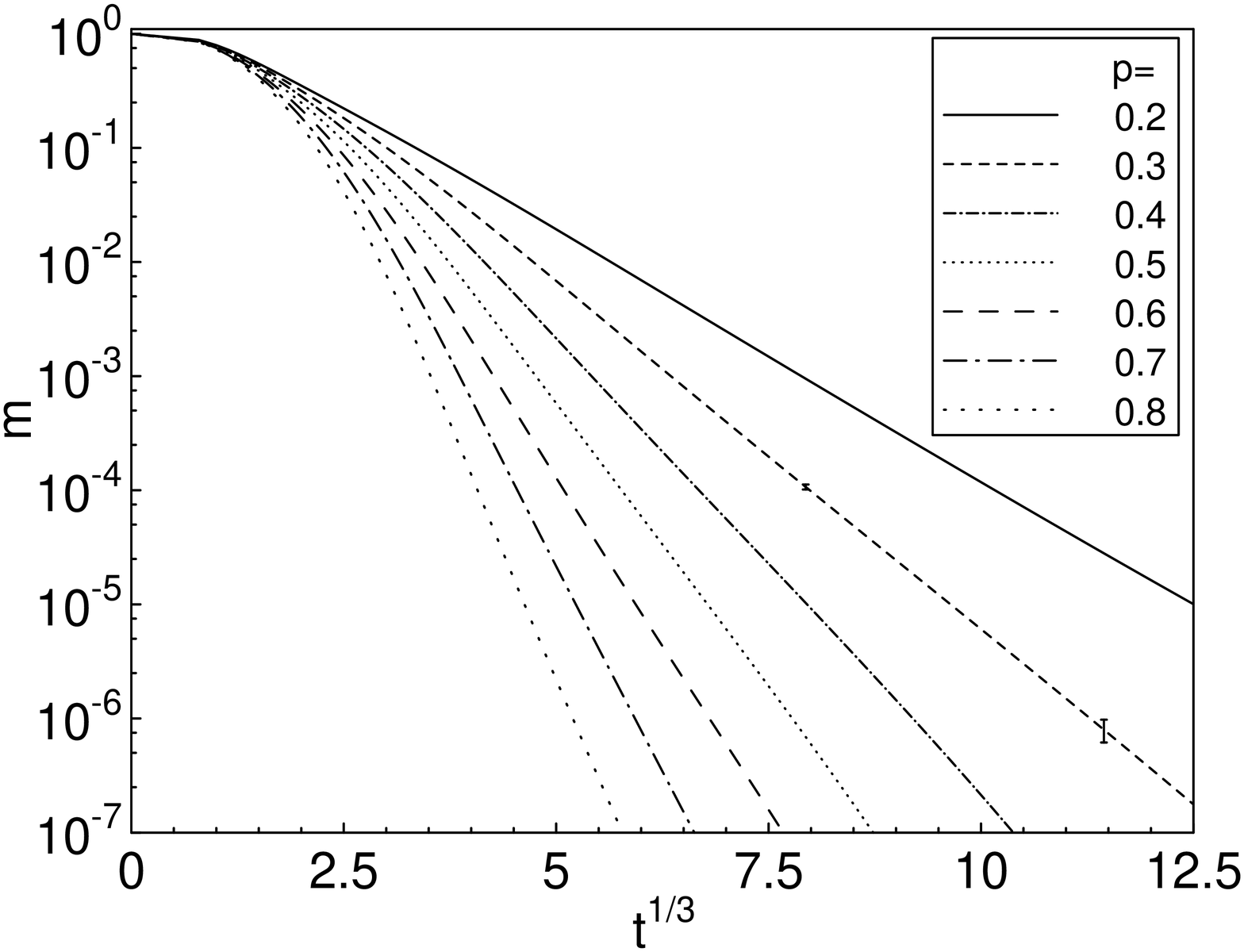}\includegraphics[width=0.5\textwidth]{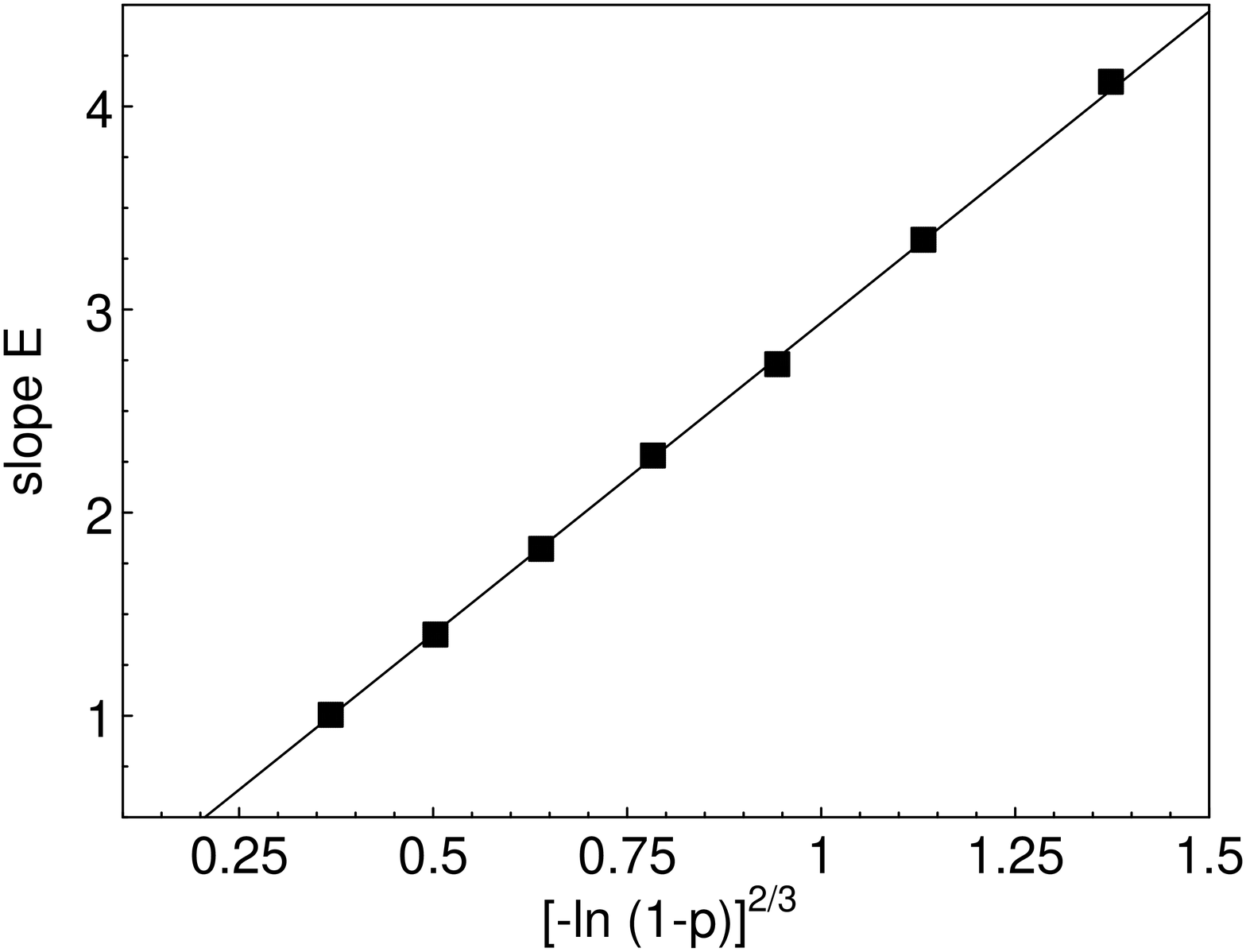}
\caption{Left: Time evolution of the magnetization at the clean critical temperature
$T_c^0=3$
   for several impurity concentrations $p$
   starting from a fully polarized state at $t_0=0$ to $t_{max}=2000$. Parameters are $L_\bot=10^5$,
   $J_0=1$, $c=0.2$. The data represent averages of 25 disorder configurations. The long-time
   behavior of $m$ follows $\ln m(t) \sim -E t^{-1/3}$. Error bars representing the standard
   deviation of $m$ are shown for $p=0.3$.
   Right: Decay constant $E$ of the stretched exponential decay of the magnetization
    as a function of the impurity concentration $p$.
     The solid line is a linear fit to $[-\ln(1-p)]^{2/3}$.}
\label{fig:evo_tc0}
\end{figure}

To study the dynamics, we have numerically integrated the equations (\ref{eq:mf}) for
$\alpha=1, J_0=1$, $c=0.2$, dilutions of $p=0.2,0.3,0.4,0.5,0.6,0.7$ and $0.8$ and system
sizes (uncorrelated direction) up to $L_\bot =10^6$ averaging over up to 25 disorder
realizations. Let us first analyze the behavior at the clean critical temperature
$T_c^0=3.0$ (at the threshold of rare region ordering). The left panel of figure
\ref{fig:evo_tc0} shows the time evolution of the total magnetization $m=(1/L_\bot)\sum_x
m_x$ at $T_c^0=3$ for several dilutions.
The figure shows that the long-time decay of the magnetization is governed by a stretched
exponential of the form $\ln m(t) =-E t^{-1/3}$. The decay constant $E$ increases with
increasing dilution. The right panel of figure \ref{fig:evo_tc0} shows $E$ as a function
of the dilution $p$. Clearly, the decay constant depends linearly on $[-\ln(1-p)]^{2/3}$.

This behavior can be understood using variational arguments similar to that of Bray
\cite{Bray88b,Bray88a} for conventional Griffiths effects. At the clean critical point,
the correlation time $\xi_t$ of an impurity-free island depends on its linear size $L_R$
via a power law $\xi_t \sim L_R^z$. Here, $z$ is the clean dynamical exponent whose value
is 2 for our infinite-range model. The rare region contribution to the time-dependent
magnetization $m(t)$ can obtained by summing over the exponential time dependencies of
the individual islands with $L_R$-dependent correlation time. Using (\ref{eq:wLR}) we
obtain to leading exponential accuracy
\begin{equation} m(t) \sim \int dL_R
~\exp \left[ L_R \ln(1-p)- t/\xi_t(L_R) \right]~. \label{eq:mf_mt}
\end{equation}
This integral can easily be evaluated variationally, i.e., within the saddle-point
method.  The main contribution to $m$ at time $t$ comes from islands of size $L_R^{SP}(t)
\sim [t/(-\ln(1-p))]^{1/3}$. The leading long-time decay of the magnetization at the
clean critical point is then given by a stretched exponential
\begin{equation}
\ln m(t) \sim - t^{1/3}~ [-\ln(1-p)]^{2/3} \label{eq:streched}
\end{equation}
in complete agreement with the simulation results presented in figure \ref{fig:evo_tc0}.

Before we continue, let us briefly comment on error bars and finite-size effects. A
detailed analysis of finite-size effects in our model was performed for the
thermodynamics in Ref.\ \cite{jpa2003}. A finite-size sample contains only a finite
number of islands of a certain size $L_R$. As long as the number of relevant islands of
size $L_R^{SP}$ is large, finite-size effects on the time evolution, Fig.\
\ref{fig:evo_tc0}, are small and governed by the central-limit theorem. With increasing
time, $L_R^{SP}$ increases; and the number of islands of this size decreases. When it
becomes of order one, strong sample-to-sample fluctuations are expected. Finally, when
$L_R^{SP}$ becomes larger than the largest island on the sample, the time evolution will
cross over from stretched exponential (\ref{eq:streched}) to simple exponential decay.
The data in Fig.\ \ref{fig:evo_tc0} show that finite-size effects do not play a
significant role in the parameter range shown. Sample-to-sample fluctuations (as
represented by the standard deviation of the magnetization) remain small down to
$m=10^{-6}$. Also, no deviations from stretched exponential decay are observed even at
the smallest magnetizations shown (they do start to occur below $m\approx 10^{-8}$,
though).

\begin{figure}
\includegraphics[width=0.5\textwidth]{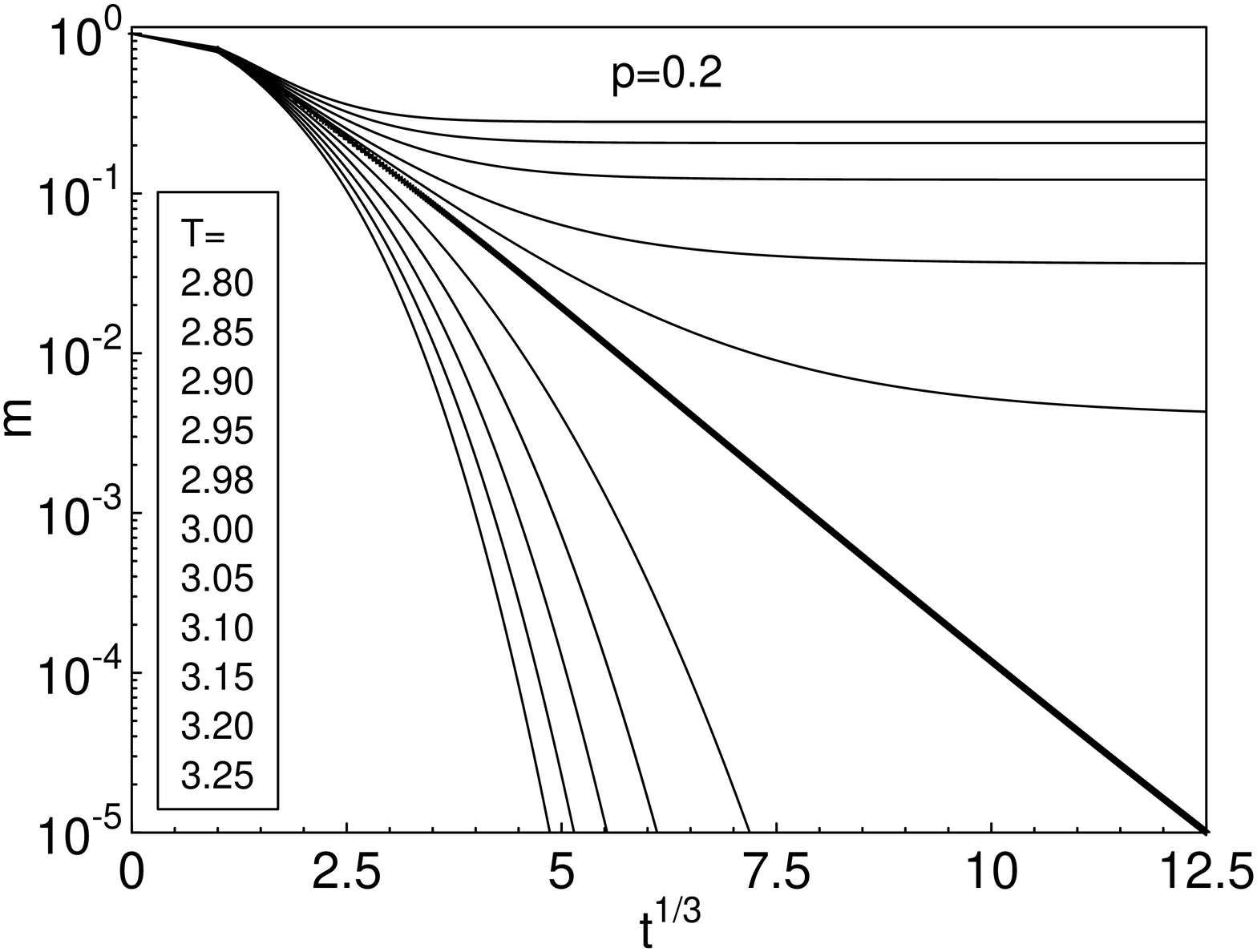}\includegraphics[width=0.5\textwidth]{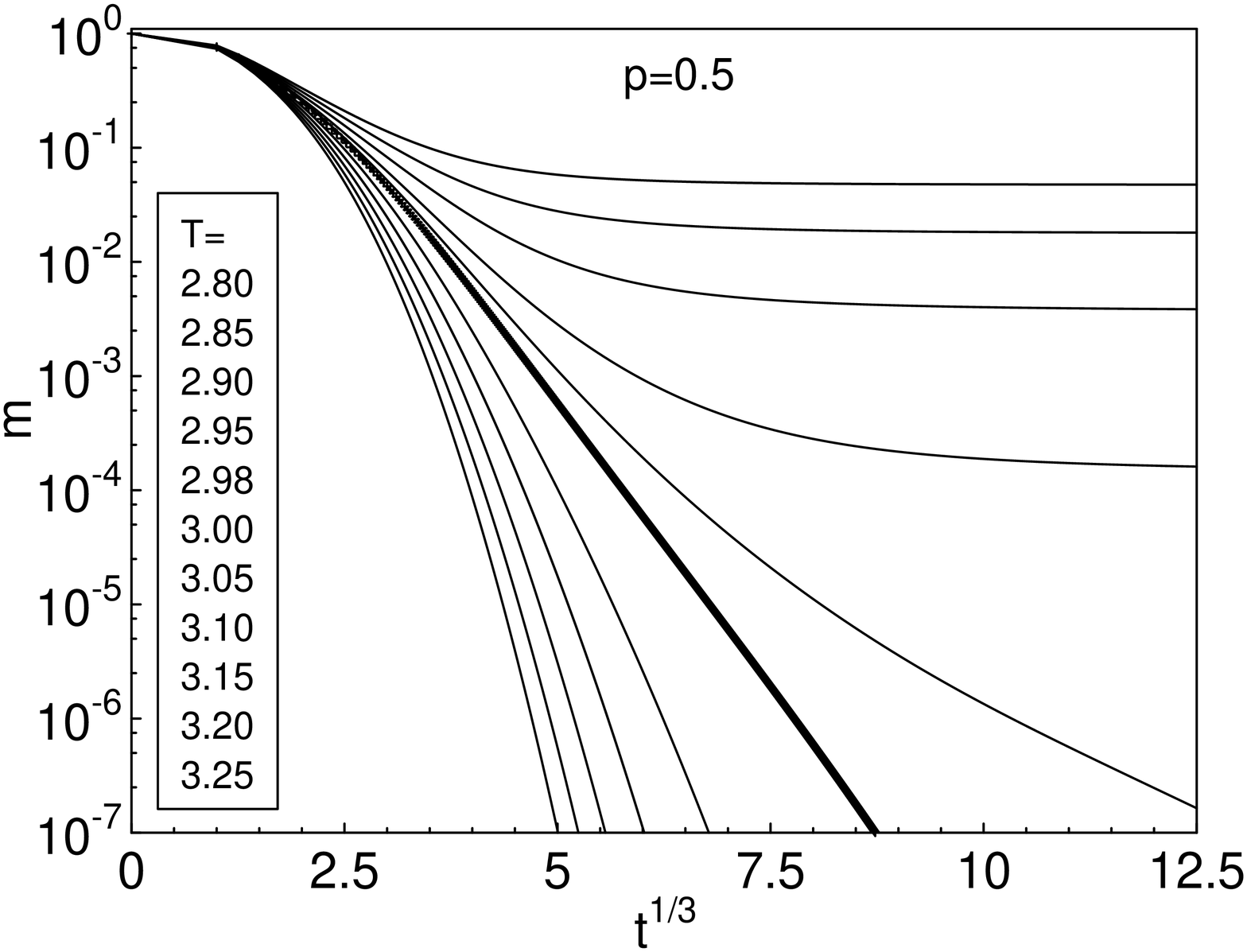}
\caption{Time evolution of the magnetization at several temperatures from $T=2.8$ to 3.25
(top to bottom) starting from a fully polarized state at $t_0=0$ to $t_{max}=2000$.
Parameters are as in figure \ref{fig:evo_tc0}. The data represent averages of 25
disorder configurations. The impurity concentrations are $p=0.2$ (left panel) and $p=0.5$
(right panel).} \label{fig:evo_off}
\end{figure}

\begin{figure}[b]
\includegraphics[width=0.5\textwidth]{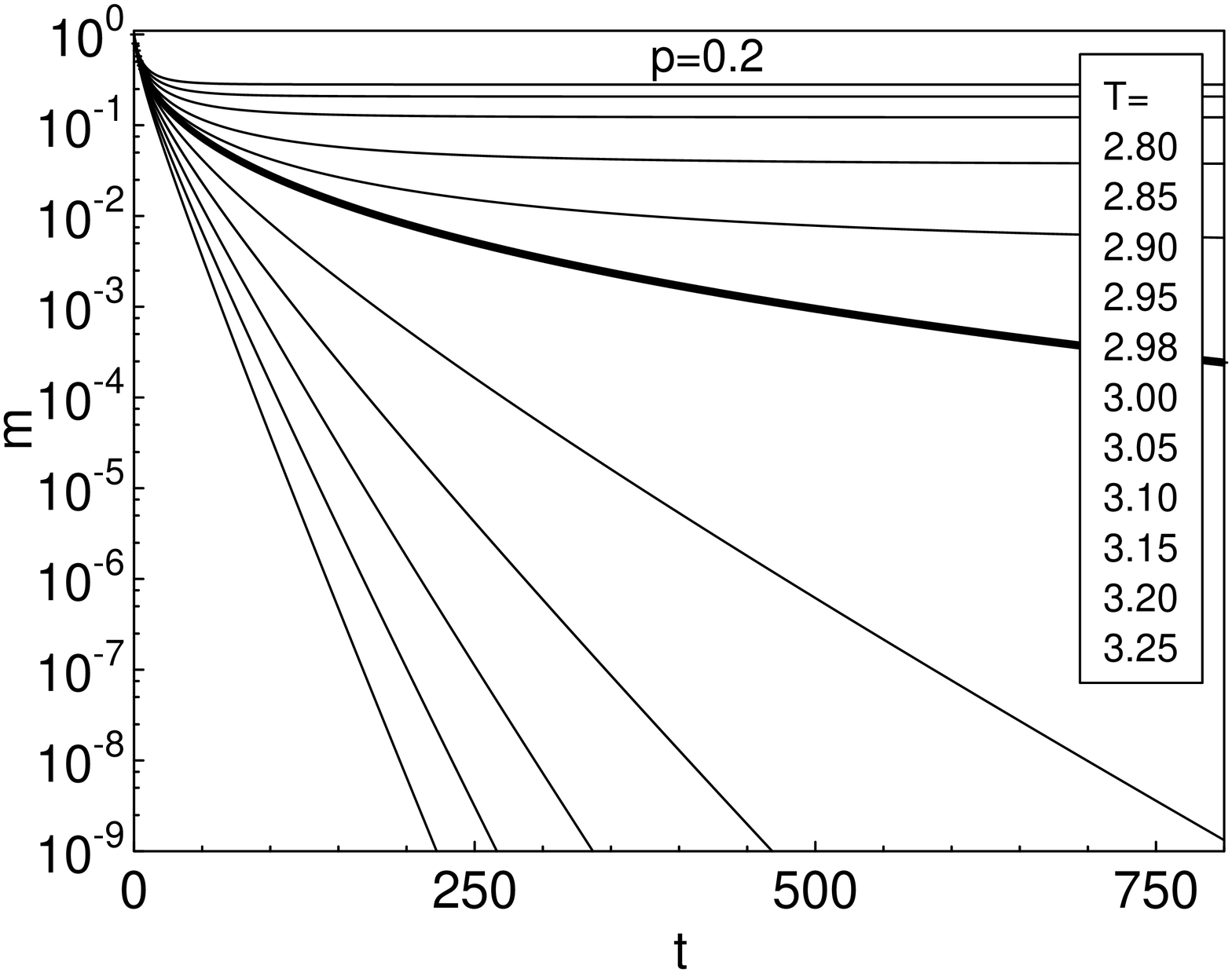}\includegraphics[width=0.5\textwidth]{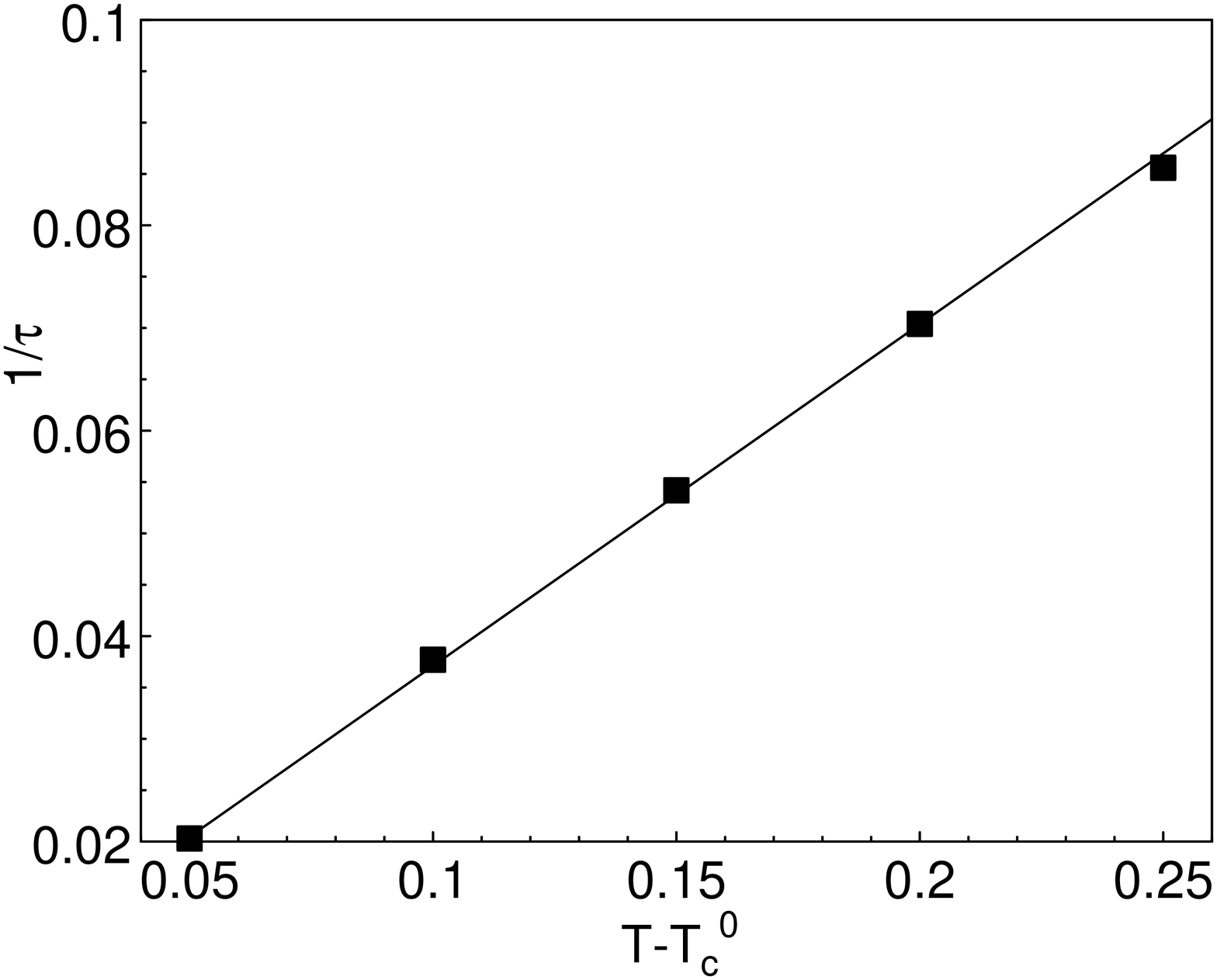}
\caption{Left: Replot of the the $p=0.2$ data from figure \ref{fig:evo_off}. Right:
Inverse decay time $1/\tau$ as a function of $T-T_c^0$ for the curves above $T_c^0$.}
 \label{fig:above_tc0}
\end{figure}

We now turn to the dynamic behavior for temperatures close to but not at the clean
critical temperature $T_c^0$. Figure \ref{fig:evo_off} shows the time evolution of the
magnetization for several temperatures from $T=2.8$ to 3.25.
At early times, all curves follow the stretched exponential found at $T_c^0$.  After
reaching a (temperature-dependent) crossover time $t_x$, the behavior qualitatively
changes. For temperatures above the clean critical temperature, the magnetization decays
faster than the stretched exponential. In fact, the asymptotic decay is exponential $\ln
m \sim t/\tau(T)$, as can be seen from a log-linear replot of the $p=0.2$ data in the
left panel of figure \ref{fig:above_tc0}.
The dependence of relaxation time $\tau(T)$ on the temperature $T$ is shown in the right
panel of figure \ref{fig:above_tc0}. Clearly, 1/$\tau$ varies linearly with $T-T_c^0$ in
good approximation. This behavior can be understood as follows. For $T>T_c^0$ the
correlation time of the largest islands does not diverge, but it is cut off by the
distance from the clean critical point, $\xi_t \sim (T-T_c^0)^{-z\nu}$. For our
infinite-range model, $z\nu =1$. The large islands with this correlation time dominate
the variational integral (\ref{eq:mf_mt}) for $m(t)$. This leads to a simple exponential
decay with an inverse decay time that depends linearly on $T-T_c^0$.

\begin{figure}
\centerline{\includegraphics[width=0.5\textwidth]{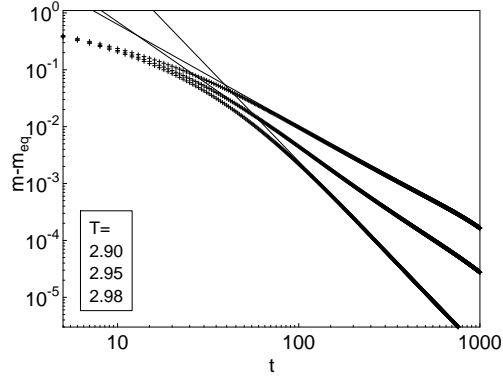}}
\caption{Double-logarithmic plot of the approach of the magnetization to the equilibrium
value in the tail of the smeared transition for $T=2.90$, 2.95 and 2.98 (top to bottom).
The impurity concentration is $p=0.5$, the other parameters are as in figure
\ref{fig:evo_tc0}. The straight lines are fits to power laws yielding exponents of 1.78,
2.20, and 3.31, respectively.} \label{fig:power_law}
\end{figure}

The more interesting case are temperatures below the clean critical temperature $T_c^0$,
i.e., the tail region of the smeared transition. Here, the total magnetization approaches
an exponentially small but non-zero value in the long-time limit, in agreement with
(\ref{eq:m-dilute}) and Ref. \cite{jpa2003}. The approach to this stationary value is
very slow. Fig.\ \ref{fig:power_law} shows $m(t)-m(\infty)$ vs. $t$ for $p=0.5$ and three
different temperatures.
The data show that the asymptotic decay is governed by a power law with a non-universal,
temperature-dependent exponent. To understand this, we note that below $T_c^0$ the island
correlation time has a divergence of the form $\xi_t(T, L_R) \sim | T_c^0 -T - A L_R
^{-2}|^{-1}$. If this is inserted into the variational integral (\ref{eq:mf_mt}), the
main contribution comes from (finite-size) islands with $L_R^{-2} \approx (T_c^0-T)/A$
because they have diverging correlation time. Carrying out the integral leads to a
power-law for $m(t)-m(\infty)$.

\begin{figure}
\centerline{\includegraphics[width=0.5\textwidth]{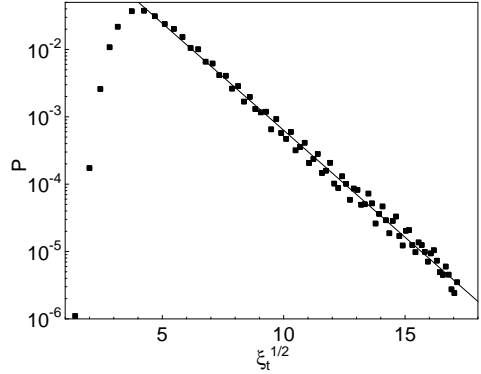}}
\caption{Probability
distribution of the local relaxation times at $T_c^0$ for $p=0.3$, determined from 10
samples of size $L_\bot=10^6$. The other parameters are as in figure \ref{fig:evo_tc0}.
The solid line is a fit of the large-$\xi_t$ tail to $\ln P \sim \xi_t^{-1/2}$.}
\label{fig:distrib}
\end{figure}

The distribution of island sizes leads to a corresponding distribution of local
correlation times $\xi_t(x)$. Within our infinite-range model, they can be determined by
analyzing the time evolution of the local magnetization via $\ln m_x \sim -t/\xi_t(x)$.
Figure \ref{fig:distrib} shows the resulting probability distribution $P(\xi_t)$ of the
local correlation times at the clean critical point $T_c^0$ for $p=0.3$ as determined
from 10 samples of $L_\bot=10^6$ sites.
The behavior of the large-$\xi_t$ tail of the probability distribution follows
from combining eq.\ (\ref{eq:wLR}) with $\xi_t \sim L^2$ leading to
$\ln P \sim \xi_t^{-1/2}$.

\section{Dynamics from general scaling arguments}
\label{sec:dyn}

In this section, we use general scaling arguments to determine the dynamical behavior at
the smeared transition of our disordered magnet (\ref{eq:Hamiltonian}) beyond mean-field
theory. The interesting physics in the tail of the smeared transition is local because
the islands are effectively decoupled from each other, and the spatial correlation length
remains finite. An appropriate quantity to study the rare region dynamics is therefore
the spin autocorrelation function
\begin{equation}
C(t) = \frac 1 V \sum_{\bf x} ~\langle S_{\bf x}(t) S_{\bf x}(0)\rangle~.
\end{equation}
To determine the leading behavior of $C(t)$ in the tail of the smeared transition
we generalize the variational extremal statistics calculation from section \ref{sec:numerics}
to the non-mean-field case.

According to finite-size scaling \cite{barber}, the behavior of the correlation time
$\xi_t$ of a single rare region of size $L_R$ in the vicinity of the clean bulk critical
point can be modelled by (for $T<T_c^0$, i.e., $r<0$)
\begin{equation}
\xi_t(r, L_R) \sim L_R^{(z\nu - \tilde z\tilde \nu)/\nu}
   \left|  r + A L_R ^{-1/\nu}\right|^{-\tilde z \tilde \nu}~. \label{eq:xit}
\end{equation}
Here, $\nu$ and $z$ are the correlation length and dynamical exponents of a $d$-dimensional system, and $\tilde \nu$ and
$\tilde z$ are the corresponding exponents of a $d_C$-dimensional system.

Let us first consider the time evolution of the autocorrelation function $C(t)$ at the
clean critical point $T_c^0$. For $r=0$, the correlation time (\ref{eq:xit}) simplifies
to $\xi_t \sim L_R^z$. Similar to the total magnetization in the mean-field treatment above,
the rare region contribution to $C(t)$ is obtained by simply
summing over the exponential time dependencies of the individual islands with
$L_R$-dependent correlation time. Using (\ref{eq:wLR}) we obtain to exponential accuracy
\begin{equation} C(t) \sim \int dL_R
~\exp \left ( L_R^{d_\bot} \ln(1-p) -D t/L_R^z \right)
\label{eq:ct}
\end{equation}
where $D$ is a constant. This integral can easily be evaluated within the saddle-point
method.  The leading long-time decay of the autocorrelation function $C(t)$ at the clean
critical point is given by a stretched exponential,
\begin{equation}
\ln C(t) \sim -[-\ln(1-p)]^{z/(d_\bot+z)}~ t^{d_\bot/(d_\bot+z)}~. \label{eq:stretched}
\end{equation}

For $T>T_c^0$, i.e. $r>0$, the correlation time does not diverge for any $L_R$. Instead,
the correlation time of the large islands saturates at $\xi_t(r, L_R) \sim r^{-z \nu}$
for $L_R > (r/A')^{-\nu}$. The autocorrelation function $C(t)$ can again be evaluated as
an integral over all island contributions. For intermediate times $t<t_x\sim
|r|^{-(d_\bot+z)\nu}$, the autocorrelation function follows the stretched exponential
(\ref{eq:stretched}). For times larger than the crossover time $t_x$, we obtain a simple
exponential decay $\ln C(t) \sim -t/\tau$ with the decay time $\tau \sim r^{-z\nu}$. Our
results for $T \ge T_c^0$ agree with those for conventional Griffiths effects
\cite{Bray88a}. This is not surprising, because above $T_c^0$, there is no difference
between the Griffiths and the smearing scenarios: In both cases, all rare regions are
locally still in the disordered phase.

This changes below the clean critical point $T_c^0$. For $r<0$, we repeat the saddle
point analysis with the full expression (\ref{eq:xit}) for the correlation length. Again,
for intermediate times $t<t_x$, the decay of the average density is given by the
stretched exponential (\ref{eq:stretched}). For times larger than the crossover time
$t_x$ the system realizes that some of the rare regions have developed static order and
contribute to a non-zero equilibrium value of the autocorrelation function $C(t)$. The
approach of $C(t)$ to this equilibrium value is dominated by finite-size islands with
$L_R \sim (-r/A)^{-\nu}$ because they have diverging correlation time.  As a result, we
obtain a power-law.
\begin{equation}
C(t) - C(\infty) \sim t^{-\psi}~. \label{eq:power}
\end{equation}
The value of $\psi$ cannot be found by our methods since it depends on the neglected
pre-exponential factors.

The behavior of the \emph{local} autocorrelation function $C({\bf x},t) = \langle S_{\bf
x}(t) S_{\bf x}(0)\rangle$ is spatially inhomogeneous because the local correlation time
depends on the size of the underlying rare region. The probability distribution
$P(\xi_t)$ of these of local (island) relaxation times can be obtained from
(\ref{eq:wLR}) and (\ref{eq:xit}). At the clean critical point $T_c^0$ we find
\begin{equation}
P(\xi_t) \sim \exp (-D\,[-\ln(1-p)]\, \xi_t^{d_\bot/z})
\end{equation}
where $D$ is a constant.

\section{Conclusions}
\label{sec:conclusions}

To summarize, we have studied the dynamic behavior of Ising magnets with planar defects.
In these systems the magnetic phase transition is smeared  because rare strongly coupled
spatial regions independently undergo the phase transition.  In the rare-region dominated tail of the
smeared transition, the dynamics is very slow. Using general scaling arguments, we have
shown that the spin autocorrelation function decays in a two stages, a stretched
exponential decay at intermediate times followed by a power-law approach to the
exponentially small stationary (equilibrium) value. We have illustrated these general
scaling results by computer simulations of an infinite-range model.

Let us briefly compare our results to the dynamics in a conventional Griffiths phase
(as produced by uncorrelated disorder) \cite{Randeria,Dhar,Dhar88,Bray88a,Bray88b}.
In both cases, rare regions dominate the long-time dynamics. In a conventional Griffiths
phase, the finite-size rare regions remain fluctuating for all temperatures above the
dirty critical point, i.e., their island correlation times \emph{remain finite}. As a result, the
autocorrelation function decays like
$\ln C(t) \sim -(\ln t)^{d/(d-1)}$ for Ising systems \cite{Randeria,Dhar,Dhar88,Bray88a,Bray88b},
and  $\ln C(t) \sim -t^{1/2}$ for Heisenberg systems \cite{Bray88a,Bray87}.
In contrast, in the tail of a \emph{smeared} transition, the effects of the rare regions are even stronger
because individual islands can undergo the phase transition independently, connected
with a \emph{divergent} island correlation time. This leads to an even slower power-law
decay of the spin autocorrelation function as shown in section \ref{sec:dyn}.

We emphasize that we have considered a purely relaxational (local) dynamics corresponding
to model A in the Hohenberg-Halperin classification \cite{hh1977}. Other dynamic
algorithms require separate investigations. For instance, model B dynamics globally
conserves the order parameter. In this context, an interesting question is: How does the
interplay of the \emph{local} thermodynamics of the rare regions and the \emph{global}
conservation law modify the dynamic behavior in the tail of the seared transition.

Finally, we note that while the results here have been derived for an Ising model with
planar defects, we expect analogous results for other disorder-smeared phase transitions.
Indeed, in the tail of  the smeared non-equilibrium phase transition of a contact process with
extended impurities, a power-law decay of the density was recently found \cite{contact_pre}.

\ack We acknowledge support from the University of Missouri Research Board and from the
NSF under grants No. DMR-0339147 and PHY99-07949. Parts of this work have been performed
at the Aspen Center for Physics and the Kavli Institute for Theoretical Physics.

\end{document}